# Hall Effects on Casson Fluid Flow along a Vertical Plate

Mst. Sonia Akter[1, a)], Mohammad Rafiqul Islam[1, b)], Md. Tusher Mollah[1, c)] and Md. Mahmud Alam[2, d)]

[1]*Department of Mathematics, Bangabandhu Sheikh Mujibur Rahman Science and Technology University Gopalganj-8100, Bangladesh*
[2]*Mathematics Discipline, Khulna University, Khulna-9208, Bangladesh*

a) sonia.math50@gmail.com
b) Corresponding author: mribsmrstu@yahoo.com
c) tusher.bsmrstu@gmail.com
d) alam_mahmud2000@yahoo.com

**Abstract.** The Hall effects on Casson fluid flow along a vertical plate has been investigated numerically. The governing equations have been derived from Navier-Stokes' equation and boundary layer approximation has been employed. By using usual transformations, the obtained non-linear coupled partial differential equations have been transformed into dimensionless governing equations. These equations have been solved by applying the explicit finite difference method. The MATLAB R2015a tool has been used for numerical simulation. The stability and convergence criteria have been analyzed. The effect of some important parameters on the primary velocity, secondary velocity, temperature and concentration distributions as well as local shear stress, Nusselt number and Sherwood number have been shown graphically.

## INTRODUCTION

Casson fluid can be defined as a shear thinning liquid which is assumed to have an infinite viscosity at zero rates of shear, a yield stress below which no flow occurs, and a zero viscosity at an infinite rate of shear. Casson fluids are found to be applicable in developing models for blood oxygenators, haemodialysers and cardiovascular system. The study of Casson fluids has ample applications in mechanical engineering, industrial engineering especially in the extraction of crude oil from petroleum products and polymer processing. Casson [1] investigated originally the validity of Casson fluid model in his studies on a flow equation for pigment-oil suspensions of printing ink type. An approximate Casson fluid model for tube flow of blood has been studied by Walawander et al. [2]. Srivastava and Saxena [3] considered the two-layered model of Casson fluid flow through stenotic blood vessels: applications to the cardiovascular system. A mathematical study of peristaltic transport of a Casson fluid has been investigated by Mernone et al. [4]. Attia and Ahmed [5] considered the hydrodynamic impulsively lid-driven flow and heat transfer of a Casson fluid. The Time-dependent pressure gradient effect on unsteady MHD Couette flow and heat transfer of a Casson fluid has been studied by Ahmed et al. [6]. Mukhopadhyay et al. [7] investigated the Casson fluid flow over an unsteady stretching surface. The Casson fluid flow and heat transfer past an exponentially porous stretching surface in presence of thermal radiation has been studied by Pramanik [8]. Kabir and Alam [9] considered the unsteady Casson fluid flow through parallel plates with hall current, joule heating and viscous dissipation. The heat and mass transfer in magnetohydrodynamic Casson fluid over an exponentially permeable stretching surface has been investigated by Raju et al. [10].

Hence our aim is to study the Hall effects on Casson fluid flow along a vertical plate. The governing equation is concerned with the Hall effects including the thermal radiation, heat source and viscous dissipation. The usual transformations have been used to obtain the non-dimensional coupled non-linear partial differential equations. The explicit finite difference technique has been used to solve the dimensionless governing equations. The obtained results have been shown graphically.

# MATHEMATICAL FORMULATION

Consider unsteady, laminar and viscous electrically conducting incompressible Casson fluid flows along a vertical semi-infinite plate at $y=0$. The fluid flow is assumed to be in the $X$-direction which is taken along the plate in the upward direction and $Y$-axis is normal to it as shown in Fig. 1. Instantaneously at time $t>0$, the plate temperature and concentration are raised to $T_w(>T_\infty)$ and $C_w(>C_\infty)$ respectively, which are thereafter sustained. Here, $T_w, C_w$ are temperature and concentration at the wall and $T_\infty, C_\infty$ designate the temperature and concentration outside the boundary layer. A uniform magnetic field **B** is imposed parallel to the $Y$ axis. Due to the consideration of the thermal radiation, the Rosseland approximation for thermal radiation, $q_r = -\dfrac{4\sigma^*}{3k^*}\left(\dfrac{\partial T^4}{\partial y}\right)$ is introduced, which is thereafter takes the form, $q_r = -\dfrac{16\sigma^*}{3k^*}T_\infty^3\dfrac{\partial T}{\partial y}$ in association with the Taylor series for $T^4$ about $T_\infty$ implies $T^4 \cong 4T_\infty^3 T - 3T_\infty^4$. For the case of Hall effect, a $Z$-component for the velocity is expected to arise. Thus the fluid velocity vector is, $\mathbf{q} = u\hat{i} + v\hat{j} + w\hat{k}$.

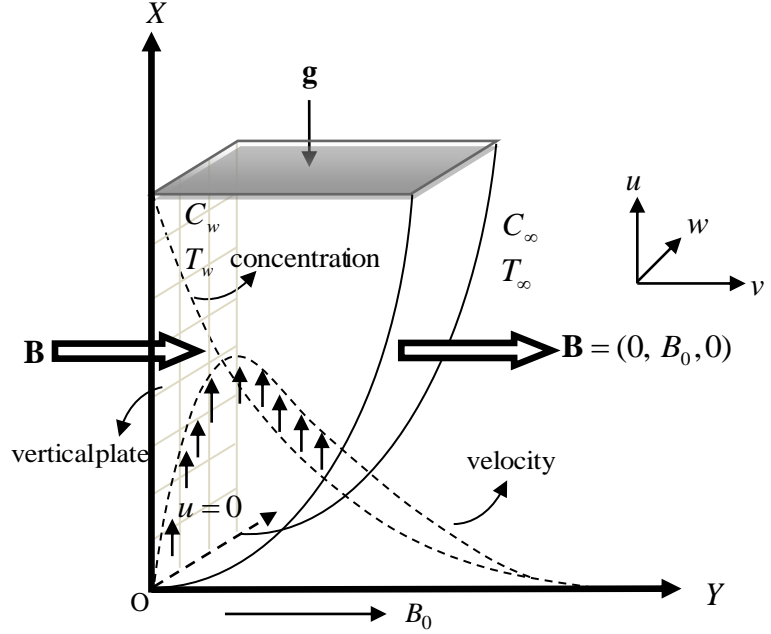

**Figure 1:** Physical Configuration and Coordinates System

The non-dimensional variables that have been used in the governing equations can be written as follows:

$$X = \frac{xU_0}{\upsilon},\ Y = \frac{yU_0}{\upsilon},\ U = \frac{u}{U_0},\ V = \frac{v}{U_0},\ W = \frac{w}{U_0},\ \tau = \frac{tU_0^2}{\upsilon},\ \theta = \frac{(T-T_\infty)}{(T_w-T_\infty)},\ \phi = \frac{(C-C_\infty)}{(C_w-C_\infty)}$$

Using these above dimensionless variables, the obtained dimensionless equations are given as follows:

$$\frac{\partial U}{\partial X} + \frac{\partial V}{\partial Y} = 0 \qquad (1)$$

$$\frac{\partial U}{\partial \tau} + U\frac{\partial U}{\partial X} + V\frac{\partial U}{\partial Y} = \left(1+\frac{1}{\beta}\right)\frac{\partial^2 U}{\partial Y^2} + G_r + G_r^* - \frac{M}{(1+m^2)}(U+mW) \qquad (2)$$

$$\frac{\partial W}{\partial \tau} + U\frac{\partial W}{\partial X} + V\frac{\partial W}{\partial Y} = \left(1+\frac{1}{\beta}\right)\frac{\partial^2 W}{\partial Y^2} - \frac{M}{(1+m^2)}(W-mU) \qquad (3)$$

$$\frac{\partial \theta}{\partial \tau} + U\frac{\partial \theta}{\partial X} + V\frac{\partial \theta}{\partial Y} = \left(\frac{Q_r+1}{P_r}\right)\frac{\partial^2 \theta}{\partial Y^2} + E_c\left(1+\frac{1}{\beta}\right)\left[\left(\frac{\partial U}{\partial Y}\right)^2 + \left(\frac{\partial W}{\partial Y}\right)^2\right] + \frac{J_h}{(1+m^2)}\left(U^2+W^2\right) - \tilde{Q}\theta \qquad (4)$$

$$\frac{\partial \phi}{\partial \tau} + U\frac{\partial \phi}{\partial X} + V\frac{\partial \phi}{\partial Y} = \frac{1}{S_c}\frac{\partial^2 \phi}{\partial Y^2} + S_r\frac{\partial^2 \theta}{\partial Y^2} \qquad (5)$$

Where, $\beta = \dfrac{\mu_b\sqrt{2\pi}}{p_y}$ is the Casson fluid parameter.

The corresponding dimensionless initial and boundary conditions can be written as follows:
$\tau > 0,\ U = 0,\ W = 0,\ \theta = 1,\ \phi = 1$ at $Y=0$ and $U \to 0,\ W \to 0,\ \theta \to 0,\ \phi \to 0$ at $Y \to \infty$

The non-dimensional parameters are given as follows:

Grashof number, $G_r = \dfrac{g\beta_T (T - T_\infty)\upsilon}{U_0^3}$; Modified Grashof number, $G_r^* = \dfrac{g\beta_C (C - C_\infty)\upsilon}{U_0^3}$; Magnetic parameter, $M = \dfrac{\sigma B_0^2 \upsilon}{\rho U_0^2}$; Prandtl number, $P_r = \dfrac{\rho C_p \nu}{k}$; Eckert number, $E_c = \dfrac{U_0^2}{C_p (T_w - T_\infty)}$; Radiative parameter, $Q_r = \dfrac{16\sigma^* T_\infty^3}{3k^* k}$; Schmidt number, $S_c = \dfrac{\nu}{D_m}$; Heat Source parameter, $\tilde{Q} = \dfrac{Q\nu}{\rho C_p U_0^2}$; Soret number, $S_r = \dfrac{D_T}{\nu} \dfrac{(T_w - T_\infty)}{(C_w - C_\infty)}$ and Joule Heating parameter, $J_h = M E_c$.

## SHEAR STRESS, NUSSELT NUMBER AND SHERWOOD NUMBER

From the velocity field, the effects of various parameters on the shear stress have been computed. The following equations represent the local shear stress at the plate. For primary velocity, the local shear stress in $X$-direction i.e. the local primary shear stress is, $\tau_{LX} = \mu \left(\dfrac{\partial U}{\partial Y}\right)_{Y=0}$. For secondary velocity, the local shear stress in $Z$-direction i.e. the local secondary shear stress is, $\tau_{LZ} = \mu \left(\dfrac{\partial W}{\partial Y}\right)_{Y=0}$. From the temperature field, the effects of various parameters on the Nusselt number have been investigated. The local Nusselt number is, $Nu_L = -\mu \left(\dfrac{\partial \theta}{\partial Y}\right)_{Y=0}$. From the concentration field, the effects of various parameters on the Sherwood number have been calculated. The local Sherwood number is, $Sh_L = -\mu \left(\dfrac{\partial \phi}{\partial Y}\right)_{Y=0}$.

## NUMERICAL TECHNIQUE

To solve the governing non-linear coupled dimensionless partial differential equations (1) to (5) with the associated initial and boundary conditions, the explicit finite difference method has been used. To obtain the difference equations in the region of the flow is divided into a grid or mesh of lines parallel to $X$ and $Y$ axes where $X$-axis is taken along the plate and $Y$-axis is normal to the plate as shown in Fig.1.

It is assumed that the maximum length of the boundary layer is $X_{\max} = 60$ i.e. $X$ varies from $0$ to $60$ and the number of grid spacing in $X$ direction is $m = 60$, hence the constant mesh size along $X$ axis becomes $\Delta X = 1.0 (0 \leq X \leq 60)$ and $Y_{\max} = 20$ i.e. $Y$ varies from $0$ to $20$ and the number of grid spacing in $Y$ direction is $n = 60$, hence the constant mesh size along $Y$ axis becomes $\Delta Y = 0.33 (0 \leq Y \leq 20)$ with a smaller time-step $\Delta \tau = 0.001$.

Let $U', W', \theta'$ and $\phi'$ denote the values of $U, W, \theta$ and $\phi$ at the end of a time-step respectively. Using the explicit finite difference method to the system of partial differential equations (1) to (5), the obtained appropriate finite difference equations are given as follows:

$$\dfrac{U_{i,j} - U_{i-1,j}}{\Delta X} + \dfrac{V_{i,j} - V_{i-1,j}}{\Delta Y} = 0 \qquad (6)$$

$$\dfrac{U'_{i,j} - U_{i,j}}{\Delta \tau} + U_{i,j} \dfrac{U_{i,j} - U_{i-1,j}}{\Delta X} + V_{i,j} \dfrac{U_{i,j} - U_{i,j-1}}{\Delta Y} = \left(1 + \dfrac{1}{\beta}\right)\left(\dfrac{U_{i,j+1} - 2U_{i,j} + U_{i,j-1}}{(\Delta Y)^2}\right) + G_r + G_r^* \\ - \dfrac{M}{(1+m^2)}\left(U_{i,j} + m W_{i,j}\right) \qquad (7)$$

$$\frac{W'_{i,j} - W_{i,j}}{\Delta \tau} + U_{i,j} \frac{W_{i,j} - W_{i-1,j}}{\Delta X} + V_{i,j} \frac{W_{i,j} - W_{i,j-1}}{\Delta Y} = \left(1 + \frac{1}{\beta}\right)\left(\frac{W_{i,j+1} - 2W_{i,j} + W_{i,j-1}}{(\Delta Y)^2}\right) - \frac{M}{(1+m^2)}\left(W_{i,j} - mU_{i,j}\right) \quad (8)$$

$$\frac{\theta'_{i,j} - \theta_{i,j}}{\Delta \tau} + U_{i,j} \frac{\theta_{i,j} - \theta_{i-1,j}}{\Delta X} + V_{i,j} \frac{\theta_{i,j} - \theta_{i,j-1}}{\Delta Y} = \left(\frac{Q_r + 1}{P_r}\right)\left(\frac{\theta_{i,j+1} - 2\theta_{i,j} + \theta_{i,j-1}}{(\Delta Y)^2}\right) + \frac{J_h}{(1+m^2)}\left[(U_{i,j})^2 + (W_{i,j})^2\right]$$
$$+ E_c\left(1 + \frac{1}{\beta}\right)\left[\left(\frac{U_{i,j} - U_{i,j-1}}{\Delta Y}\right)^2 + \left(\frac{W_{i,j} - W_{i,j-1}}{\Delta Y}\right)^2\right] - \tilde{Q}\theta_{i,j} \quad (9)$$

$$\frac{\phi'_{i,j} - \phi_{i,j}}{\Delta \tau} + U_{i,j} \frac{\phi_{i,j} - \phi_{i-1,j}}{\Delta X} + V_{i,j} \frac{\phi_{i,j} - \phi_{i,j-1}}{\Delta Y} = \frac{1}{S_c}\left(\frac{\phi_{i,j+1} - 2\phi_{i,j} + \phi_{i,j-1}}{(\Delta Y)^2}\right) + S_r\left(\frac{\theta_{i,j+1} - 2\theta_{i,j} + \theta_{i,j-1}}{(\Delta Y)^2}\right) \quad (10)$$

and the boundary conditions with the finite difference scheme are given as follows:
$\tau > 0$, $U_{i,L} = 0$, $W_{i,L} = 0$, $\theta_{i,L} = 1$, $\phi_{i,L} = 1$ at $L = 0$ and $U_{i,L} \to 0$, $W_{i,L} \to 0$, $\theta_{i,L} \to 0$, $\phi_{i,L} \to 0$ at $L \to \infty$

Here the subscripts $i$ and $j$ designate the grid points with $X$ and $Y$ coordinates respectively.

## STABILITY AND CONVERGENCE ANALYSIS

Since an explicit procedure is being used, the analysis will remain incomplete unless the stability and convergence of the finite difference scheme are discussed. For the constant mesh sizes, the stability criteria finally can be written as follows:

$$\frac{U\Delta\tau}{\Delta X} - \frac{|V|\Delta\tau}{\Delta Y} + \left(1 + \frac{1}{\beta}\right)\frac{\Delta\tau}{(\Delta Y)^2} + \frac{M\Delta\tau}{2(1+m^2)} \leq 1, \quad \frac{U\Delta\tau}{\Delta X} - \frac{|V|\Delta\tau}{\Delta Y} + \frac{\Delta\tau}{(\Delta Y)^2}\left(\frac{Q_r + 1}{P_r}\right) + \frac{\tilde{Q}}{2} \leq 1 \text{ and } \frac{U\Delta\tau}{\Delta X} - \frac{|V|\Delta\tau}{\Delta Y} + \frac{\Delta\tau}{S_c(\Delta Y)^2} \leq 1$$

Using $\Delta Y = 0.33$, $\Delta\tau = 0.001$ and the initial condition, the above equations gives $S_c \geq 0.009$, $P_r \geq -0.013$, $\tilde{Q} \leq 10 (\tilde{Q} \neq 2)$, $Q_r \leq 5$ and $M \leq 1983$ where, $\beta > 0$, $0 < m \leq 0.1$ and $E_c = 0.10$.

## RESULTS AND DISCUSSION

Due to examine the physical condition of the developed mathematical model, the steady-state numerical values have been computed for the non-dimensional primary velocity $(U)$, secondary velocity $(W)$, temperature $(\theta)$ and concentration $(\phi)$ distributions within the boundary layer. The steady-state solution has been obtained at the dimensionless time $\tau = 15$. The effect of Magnetic parameter $(M)$ and Hall parameter $(m)$ on primary velocity $(U)$, secondary velocity $(W)$, temperature $(\theta)$ and concentration $(\phi)$ distributions as well as local primary shear stress $(\tau_{LX})$, local secondary shear stress $(\tau_{LZ})$, local Nusselt number $(Nu_L)$ and local Sherwood number $(Sh_L)$ are discussed. For brevity, the effect of other parameters such as Grashof number $(G_r)$, modified Grashof number $(G_r^*)$, Prandtl number $(P_r)$, Eckert number $(E_c)$, Radiative parameter $(Q_r)$, Schmidt number $(S_c)$, Heat Source parameter $(\tilde{Q})$, Soret number $(S_r)$, Joule Heating parameter $(J_h)$ and Casson fluid parameter $(\beta)$ are not shown.

**Mesh sensitivity test:** To obtain the appropriate mesh space for $m$ and $n$, the computations have been carried out for three different mesh spaces such as $(m,n) = (60, 60), (80, 80), (100, 100)$ as shown in Fig. 2. The curves are smooth for all mesh spaces and shows a negligible changes among the curves. Thus the mesh size $(m,n) = (60, 60)$ has been considered.

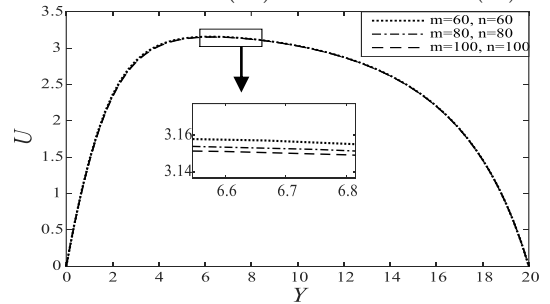

**Figure 2.** Mesh Sensitivity for Primary Velocity

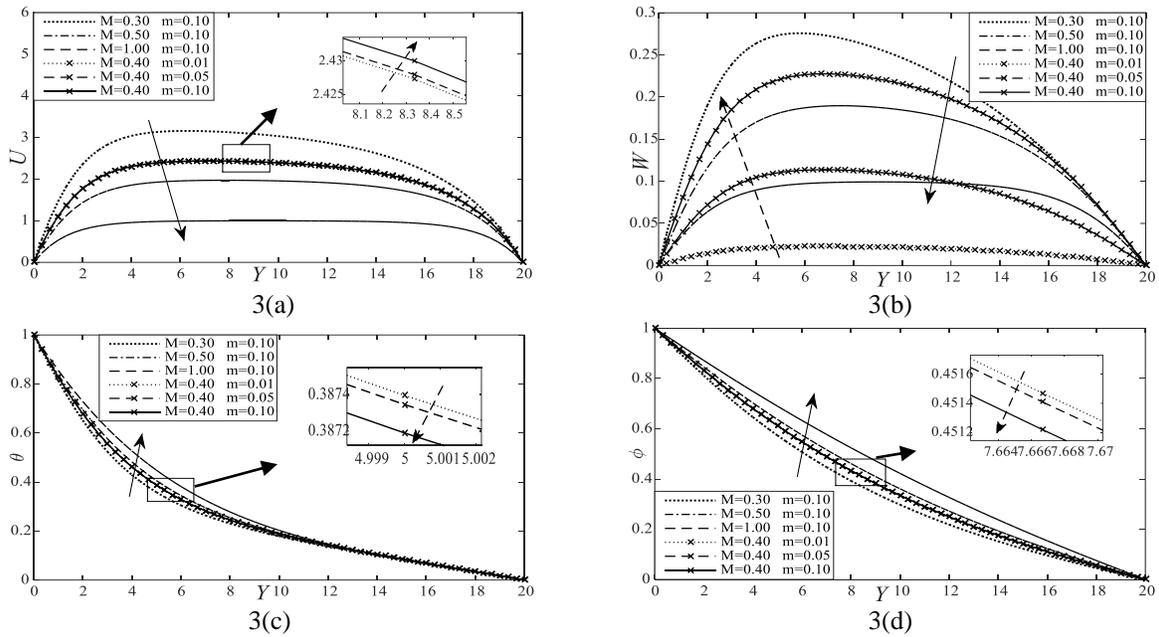

**Figure 3.** Effect of Magnetic parameter $(M)$ and Hall parameter $(m)$ on (a) Primary velocity; (b) Secondary velocity; (c) Temperature and (d) Concentration distributions; where, $G_r = 0.50$, $G_r^* = 0.50$, $E_c = 0.01$, $P_r = 0.30$, $Q_r = 0.05$, $S_c = 0.09$, $S_r = 0.09$, $\tilde{Q} = 0.05$ and $\beta = 1.00$ at time $\tau = 15$ (Steady State)

Fig. 3 shows that the primary and secondary velocities both decreases with the increase of $(M)$ while the temperature and concentration distributions both increases with the increase of $(M)$. On the other hand, the primary and secondary velocities both increases with the increase of $(m)$ while the temperature and concentration distributions both decreases with the increase of $(m)$.

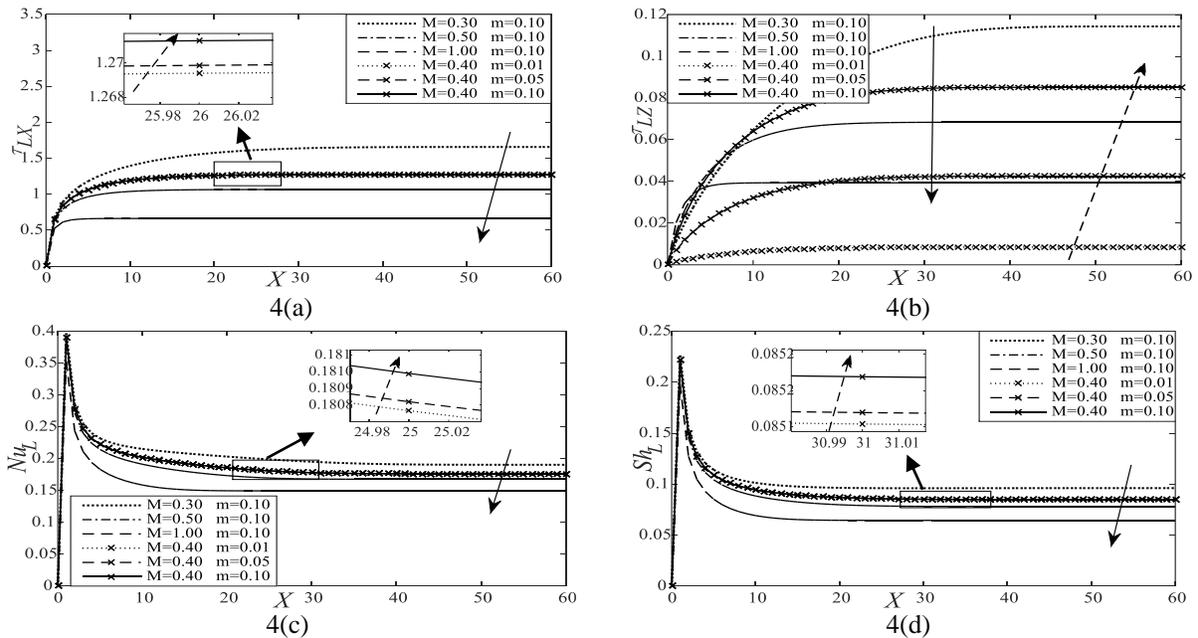

**Figure 4.** Effect of Magnetic parameter $(M)$ and Hall parameter $(m)$ on (a) local primary shear stress; (b) local secondary shear stress; (c) local Nusselt number and (d) local Sherwood number; where, $G_r = 0.50$, $G_r^* = 0.50$, $E_c = 0.01$, $P_r = 0.30$, $Q_r = 0.05$, $S_c = 0.09$, $S_r = 0.09$, $\tilde{Q} = 0.05$ and $\beta = 1.00$ at time $\tau = 15$ (Steady State)

Fig. 4 shows that the local primary shear stress, secondary shear stress, Nusselt number and Sherwood number all decreases with the increase of $(M)$. Elsewhere, the local primary shear stress, secondary shear stress, Nusselt number and Sherwood number all increase with the increase of $(m)$.

## CONCLUSION

The explicit finite difference solution for the Hall effects on unsteady viscous incompressible Casson fluid flow along a vertical plate has been investigated numerically. The numerical solutions have been found to be converged for $S_c \geq 0.009$, $P_r \geq -0.013$, $\tilde{Q} \leq 10 (\tilde{Q} \neq 2)$, $Q_r \leq 5$ and $M \leq 1983$ where, $\beta > 0$, $0 < m \leq 0.1$ and $E_c = 0.10$. The results are discussed for different values of important parameters as Magnetic parameter $(M)$ and Hall parameter $(m)$. For brevity, the effects of other parameters are not shown. Based on the results and discussion, some important findings of this investigation are mentioned as follows:

1. The primary and secondary velocities both decrease with the increase of $(M)$ while both primary and secondary velocity distributions increase with the increase of $(m)$.
2. The temperature and concentration distributions both increases with the increase of $(M)$ while both decreases with the increase of $(m)$.
3. The local primary shear stress, secondary shear stress, Nusselt number and Sherwood number all decrease with the increase of $(M)$ while all increase with the increase of $(m)$.

## ACKNOWLEDGEMENT

It is partially possible to complete the research work with the financial support of NST Fellowship under the Ministry of Science and Technology, Government of the People's Republic of Bangladesh.